\documentclass[a4paper,11pt]{article}
\usepackage{pos}
\usepackage{gensymb}

\title{Very High-energy Gamma-ray Emission from LS I +61\degree ~303 Binary}
 \ShortTitle{VHE Gamma-ray Emission from LS I +61\degree ~303 Binary}

\author*[a]{David Kieda}
\forColl{VERITAS}

\affiliation[a]{University of Utah \\
  Salt Lake City, UT, USA}

\emailAdd{david.kieda@utah.edu}

\abstract{LS I +61\degree ~303 is one of around ten gamma-ray binaries detected so far which has a spectral energy distribution dominated by MeV-GeV photons. It is located at a distance of 2 kpc and consists of a compact object (black hole or neutron star) in an eccentric orbit around a 10-15 $M_{\odot}$ Be star, with an orbital period of 26.496 days. The binary orbit modulates the emission ranging from radio to TeV energies. A second, longer, modulation period of 1667 days (the super-orbital period) has also been detected from radio to TeV observations. The VERITAS imaging atmospheric Cherenkov telescope array has been observing LS I +61\degree ~303 since 2006, and has accumulated a dataset that fully covers the entire orbit. Increased coverage of the source in the very-high-energy band is currently underway to provide more results on the modulation pattern, super-orbital period, and orbit-to-orbit variability at the highest energies. The spectral measurements at the highest energies will reveal more information about gamma-ray production/absorption mechanisms, the nature of the compact object, and the particle acceleration mechanism. Using >150 hrs of VERITAS data, we present a detailed study of the spectral energy distribution and periodic behavior of this rare gamma-ray source type at very-high energy.}

\FullConference{37$^{\rm{th}}$ International Cosmic Ray Conference (ICRC 2021)\\
		July 12th -- 23rd, 2021\\
		Online -- Berlin, Germany}

\begin{document}
\maketitle

\section*{Introduction}
 The new source class "gamma-ray binary" was coined in mid-2000s to describe high-mass binary systems with most of their radiated power emitted beyond 1 MeV . LS I +61\degree ~303 was the third gamma-ray binary detected in VHE gamma rays \cite{magic2006} after PSR B1259-63 and LS 5039. After its first detection by the Cos-B as an unidentified Galactic MeV gamma-ray emitter in 1977 \cite{cosb} to now, LS I +61\degree ~303 has been extensively studied from radio to gamma rays. It consists of a massive B0 Ve star and a compact object \cite{casares2005} and is located at a distance of 2.0 kpc \cite{distance}. The nature of the compact object remains unknown even after decades of multi-wavelength observations. The compact object could be a pulsar orbiting around a B0 main-sequence star with a circumstellar disc or it could be an accreting black hole. Neither model has successfully explained all the features of the multi-wavelength observation. However, the periodic signal with a period of 269.196 ms reported by FAST radio telescope in Jan. 2021 \cite{atel_fast} favors the pulsar scenario.

The companion star in the LS I +61\degree ~303 binary has a circumstellar disc. The shape and the size of the disc is unclear. The binary exhibits variability in its emission from radio to TeV due to orbital modulation of the emission or absorption processes \cite{2006Sci...312.1771A}. The orbital period is determined to be 26.5 days based on the period radio outbursts \cite{gregory2002}. It was also discovered in the radio wavelength that the amplitude of the radio outburst is modulated with a periodicity of 1667 days providing an evidence for the superorbital period \cite{gregory2002}. A second period of 26.94 days has also been detected in radio data \cite{massi2016} and the beat of these two orbital periods corresponds to the superorbital period. 

From the multiwavelength observation, the outbursts peaks in the radio, X-ray, and TeV emission occurs at similar orbital phase, $\phi_{orb}$, of 0.5 to 0.8 ~ \cite{zimmermann2015, casares2005, magic2006, veritas2011} while the GeV peak is anti-correlated with a shifted peak at ~ 0.25 ~ \cite{fermi2013}. The GeV spectrum is described by a power-law with an exponential cut off at 3.9 GeV  \cite{fermi2013}. It was assumed that the periastron passage occurs at an orbital phase, $\phi_{orb}$, of 0.23 and apastron passage occurs at $\phi_{orb}$ of 0.73 ~ \cite{casares2005}. However, the study by Kravtsov et. al provides the new constraints on the orbital parameter of the binary system \cite{kravtsov2020}. The orbit is close to circular with a very small eccentricity ($e<0.15$) \cite{kravtsov2020}. The periastron passage is reported at 0.6 \cite{kravtsov2020}, significantly different from the previous assumptions but corresponds to the correlated peaks in radio, X-ray and TeV around 0.6.

In this study, we report the analysis of > 150 hrs of data collected with VERITAS gamma-ray instruments. 

\section{Analysis}
The Very Energetic Radiation Imaging Telescope Array System (VERITAS) is located at the Fred Lawrence Whipple Observatory, Arizona and consists of 4 IACTs, each 12 m diameter. Each IACT is quipped with the camera system with 499 photomultiplier tubes to detect the signals from extensive air shower particles. The VERITAS gamma-ray instruments can detect $1\%$ Crab flux in less than 25 hrs and is sensitive to gamma rays in the range of 100 GeV to > 30 TeV. The VERITAS observatory started collecting data from the direction of LS I +61\degree ~303 starting 2007 after its discovery by MAGIC \cite{magic2006}. 

\begin{figure}
    \includegraphics[width=0.99\textwidth]{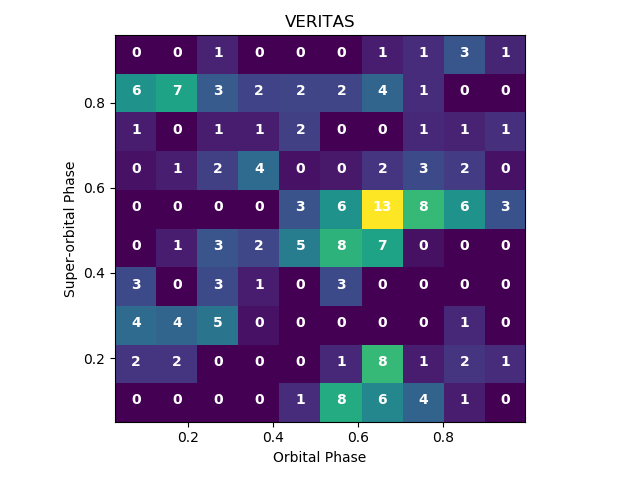}
    \caption{\textbf{Total observation of LS I +61\degree ~303 by VERITAS observatory.} Bin count is in hr. Each box indicates the total hours of A weather data observed by VERITAS in different orbital and superorbital phase bins from 2008 to 2021.}
 \label{fig:observation_hr}
\end{figure}

From 2008 to 2021, there has been about 209 hr A weather (best weather) data as shown in Fig. \ref{fig:observation_hr}. In 2009, there was a major upgrade in VERITAS, i.e relocation of one of the telescopes to achieve a more symmetrical position in the array. The data used in this study is  after the upgrade, from Oct. 2009 to Jan. 2021, covering more than a decade. The data was collected via standard VERITAS technique for point sources using wobble mode with a 0.5 ~offset. After removing the weather and hardware problems, the total good quality livetime used in the study is 164 hrs. The standard VERITAS analysis for point sources is performed with the help of an internal VERITAS software called Eventdisplay \cite{eventdisplay}. The Boosted Decision Tree technique \cite{bdt} is used at an energy threshold > 300 GeV. The background is calculated via Ring Background method in which the OFF region for the background estimation is defined as a ring around the ON region or the source region. For the Ring Background, the source exclusion radius used is 0.6 ~ and the background/source area ratio is 20. An exclusion region of 0.2 ~ is applied to the stars brighter than 7.5 mag. The significance of the gamma-ray emission from source above the background is calculated via the Li and Ma significance formula \cite{lima}.

As seen in Fig. \ref{fig:observation_hr}, the hours of observation during the flaring states ($\phi_{orb}$ = 0.5 to 0.8) are significantly more than the quiescent states ($\phi_{orb}$ = 0.8 to 0.5). In the 2014/2015 observing season, the highest ever TeV fluxes (> 30\% of the Crab Nebula) for the LS I +61\degree ~303 were recorded \cite{flares}. To date, the season consists of the highest LS I +61\degree ~303 TeV flux observed with VERITAS and a detailed study of this bright flare is provided in \cite{flares}. The entire data set is further divided into 10 orbital bins each with a width of 0.1 ~by using the starting $\phi_{orb}$ at MJD 43366.775 and an orbital period of 26.496 days. For the study in superorbital phase bins, the data is divided into 10  phase bins by using a superorbital period of 1667 days. 

\section{Results}

\begin{figure}
\includegraphics[width=0.5\textwidth]{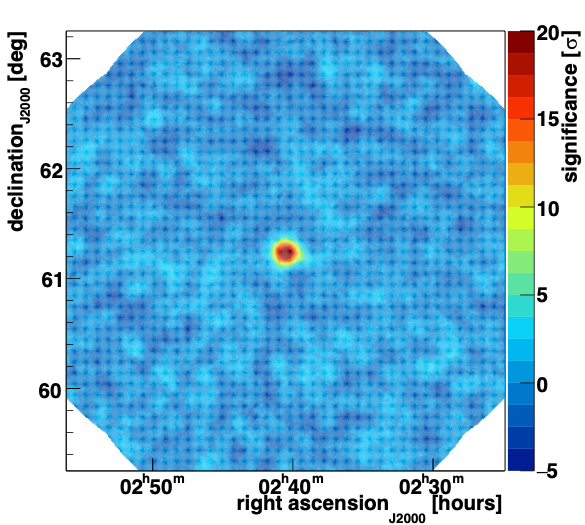}
\includegraphics[width=0.45\textwidth]{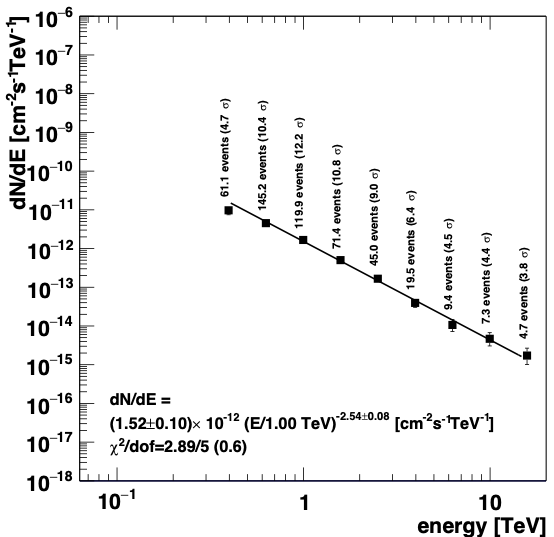}
\caption{ \label{fig:flaring}
\textbf{Left}: Significance map of the LS I +61\degree ~303 using Eventdisplay software for the $\phi_{orb}$ 0.5 to 0.8. 
\textbf{Right}: Differential fluxes above 300 GeV with $\ge$ 3 $\sigma$ detection.}
\end{figure}

\begin{figure}
\includegraphics[width=0.5\textwidth]{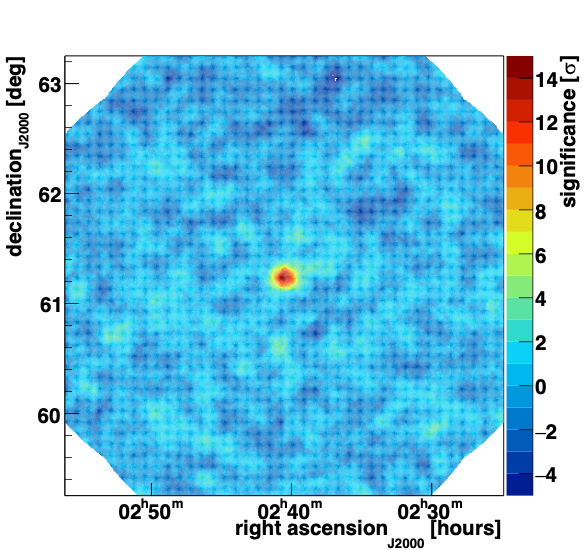}
\includegraphics[width=0.45\textwidth]{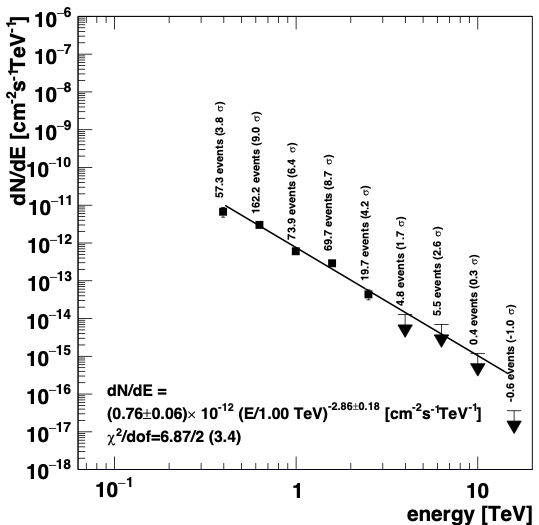}
\caption{ \label{fig:nonflaring}
\textbf{Left}: Significance map of the LS I +61\degree ~303 using Eventdisplay software for the $\phi_{orb}$ 0.8 to 0.5. 
\textbf{Right}: Differential fluxes above 300 GeV with $\ge$ 3 $\sigma$ detection. The upper limits are 99\% flux upper limit.}
\end{figure}

\begin{table*}[ht]
\centering
\begin{tabular}{c c c}
    \hline
    \textbf{$\phi_{orb}$} & \textbf{Live time (min)} & \textbf{Significance}\\   
   \hline 
      0 - 0.1 & 761.87 & 5.0\\
      0.1 - 0.2 & 760.87& 4.3\\
      0.2 - 0.3 & 738.50& 3.6\\
      0.3 - 0.4 & 760.63& 5.9\\
      0.4 - 0.5 & 759.08& 7.1\\
      0.5 - 0.6 & 1123.60& 11.7\\
      0.6 - 0.7 & 2090.47& 31.7\\
      0.7 - 0.8 & 1330.12& 9.1\\
      0.8 - 0.9 & 920.55& 6.6\\
      0.9 - 1.0 & 555.83& 4.2\\
      0.5 - 0.8 & 3286.27& 20.4\\
      0.6 - 0.5 & 5257.33& 13.9\\
      0 - 1.0 & 9801.52 & 33.0\\
    \hline  
\end{tabular}
\captionsetup{width=.9\linewidth}
\caption{\textbf{Livetime and Significance in various orbital bins}}
\label{tab:orbital_bins}
\end{table*}

The gamma-ray emission from the LS I +61\degree ~303 is detected with 33 $\sigma$ above the energy threshold of 260 GeV and using 163 hours of observation. The spectral energy distribution is described by a power law spectrum,
\begin{equation}
\label{eq:pl}
\dfrac{dN}{dE}=N_{0}\left(\dfrac{E}{ E_{0}}\right)^{\Gamma},
\end{equation}
where, the flux normalization $N_{0}$ is $(1.35\pm 0.05) \times 10^{-12} \rm \ TeV^{-1}cm^{-2}s^{-1}$, $E_{0}$ is 1 TeV and the spectral index $\Gamma$ is $-2.55 \pm 0.05$.

The live time and significance in 10 $\phi_{orb}$ bins each with 0.1 ~width are listed in Table \ref{tab:orbital_bins}. In the $\phi_{orb}$ 0.5 to 0.8, it is detected with 20.4 $\sigma$ above the energy threshold of 260 GeV and using 54.8 hours of observation  as shown in the Fig.\ref{fig:flaring} and in the $\phi_{orb}$ 0.8 to 0.5, it is detected with 13.9 $\sigma$ using 87.6 hours of observation as shown in the Fig.\ref{fig:nonflaring}. The spectral energy distribution of the $\phi_{orb}$ of (0.5 - 0.8) and (0.8 - 0.5) are shown in the Fig.\ref{fig:flaring} and Fig.\ref{fig:nonflaring} respectively. In the quieter (0.8 - 0.5) bin, the spectrum is slightly softer and there is a hint of cutoff in the energy spectrum above 10 TeV. 

\begin{figure}
    \includegraphics[width=0.99\textwidth]{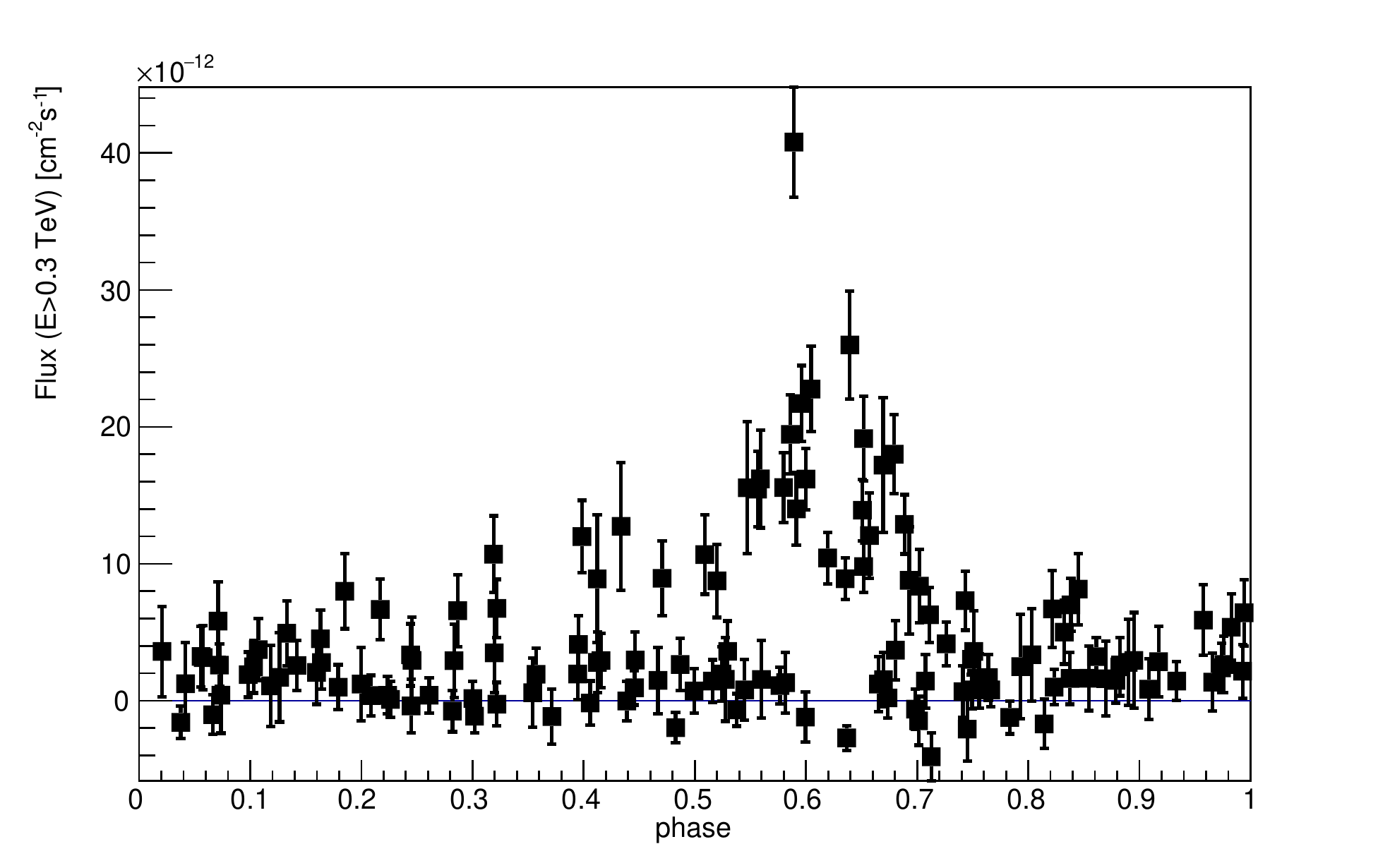}
    \caption{Nightly orbital phase binned light curve of LS I +61\degree ~303 with 163 hours of VERITAS data. The starting MJD used is 43366.8 days and the orbital period used is 26.496 days.}
 \label{fig:lightcurve}
\end{figure}

LS I +61\degree ~303 is significantly detected in most of the bins except phase bin (0.1 - 0.2), (0.2 - 0.3) and (0.9 - 1.0). In those low significance bins, LS I +61\degree  ~303 is detected with about 4$\sigma$. The highest significance is at (0.6 to 0.7) $\phi_{orb}$ bin. The nightly light curve binned with the orbital period in Fig.\ref{fig:lightcurve} shows the flux outburst at about 0.65. For the flaring states, 0.5 to 0.8 ~$\phi_{orb}$ bins, it is detected with 20.4$\sigma$ and for the quiescent states from 0.8 to 0.5, ~it is detected with 13.9 $\sigma$. The study of the energy spectrum in these various 10 orbital bins shows no significant spectral variation. The average spectral index is about -2.6. Because of the lack of the statistics, the superorbital phase study in each $\phi_{orb}$ bin could not be performed. However, the superorbital phases are explored in the $\phi_{orb}$ bins (0 - 0.3), (0.3 - 0.5), (0.5 - 0.7) ~ and (0.7 - 1.0). Only (0.5 - 0.7) bin has enough statistics, in which there is no significant spectral variation in the superorbital phases. 

\section{Summary}
LS I +61\degree ~303 is a gamma-ray binary, the high energy photons dominate its spectral energy distribution. Using more than a decade of VERITAS data, the spectral energy distribution of the binary system follows a power law spectrum. In the $\phi_{orb}$ 0.5 ~to 0.8  ~when the source is flaring, the spectral index is -2.54$\pm$0.08. For the bin 0.8 to 0.5, the spectral index is slightly softer at -2.86$\pm$0.18. The emission peaks in the $\phi_{orb}$ 0.6 ~to 0.7.

A correlated X-ray to TeV gamma-ray emission but anti-correlated to the GeV emission has also been observed in other binary system such as LS 5039 \cite{chang2016}. However, LS I +61\degree ~303 include a more complex stellar wind structure, a circumstellar disc around the companion star in comparison to LS 5039 for which no such disc has been observed. 

According to Bednarek et al. \cite{twopopulation}, the two leptonic population can explain these characteristics in X-ray to TeV observation. Assuming that the compact object is a pulsar, interaction between the pulsar and the companion star winds creates a double shock structure within the binary system. The leptonic population is accelerated to higher energies on the shock from the side of the pulsar in comparison to the accelerated leptonic population from the side of massive star. Hence, the multi-wavelength gamma-ray emission consists of two components of gamma-ray photons, which also explain a different spectral shape, a cutoff, obtained in GeV energies \cite{fermi2013}. According to Huber et al. \cite{huber}, “a complex distribution of accelerated particles” in the massive star wind and pulsar interaction environment can explain the orbital modulation observed in all band.

A super-orbital modulation has also been observed in all wavelength. A long-term variability of 1610 days, close to 4.5 yr observed in radio, has been reported by MAGIC. In case of VERITAS, due to the observation statistics, only the $\phi_{orb}$ bin 0.5  ~to 0.7 has enough statistics for a possible search for superorbital variability. While we did not observe any spectral variation for the superorbital phases in the $\phi_{orb}$ 0.5  ~to 0.7, further studies to look for a long-term modulation are in process. Additionally, modeling of the VERITAS dataset is underway, following the procedure of Mariaud et al. \cite{mariaud}.

\section{Acknowledgement}
This research is supported by grants from the U.S. Department of Energy Office of Science, the U.S. National Science Foundation and the Smithsonian Institution, by NSERC in Canada, and by the Helmholtz Association in Germany. This research used resources provided by the Open Science Grid, which is supported by the National Science Foundation and the U.S. Department of Energy's Office of Science, and resources of the National Energy Research Scientific Computing Center (NERSC), a U.S. Department of Energy Office of Science User Facility operated under Contract No. DE-AC02-05CH11231. We acknowledge the excellent work of the technical support staff at the Fred Lawrence Whipple Observatory and at the collaborating institutions in the construction and operation of the instrument.

\bibliographystyle{plain}
\bibliography{bibliography}

\begin{thebibliography}{10}

\bibitem{veritas2011}
V.~A. {Acciari} et~al.
\newblock {VERITAS Observations of the TeV Binary LS I +61{\textdegree} 303
  During 2008-2010}.
\newblock {\em Astrophys. J. Lett.}, 738(1):3, September 2011.

\bibitem{fermi2013}
M.~Ackermann et~al.
\newblock Associating long-term gamma-ray variability with the superorbital
  period of ls i + 61◦ 303.
\newblock {\em The Astrophysical Journal Letters}, 773:L35, 08 2013.

\bibitem{magic2006}
J.~{Albert} et~al.
\newblock {Variable Very-High-Energy Gamma-Ray Emission from the Microquasar LS
  I +61 303}.
\newblock {\em Science}, 312(5781):1771--1773, June 2006.

\bibitem{2006Sci...312.1771A}
J.~{Albert} et~al.
\newblock {Variable Very-High-Energy Gamma-Ray Emission from the Microquasar LS
  I +61 303}.
\newblock {\em Science}, 312(5781):1771--1773, June 2006.

\bibitem{flares}
S.~{Archambault} et~al.
\newblock {Exceptionally Bright TeV Flares from the Binary LS I +61 303}.
\newblock {\em Astrophys. J. Lett.}, 817(1):L7, January 2016.

\bibitem{twopopulation}
W.~Bednarek.
\newblock {A model for the two component gamma‐ray spectra observed from the
  {\ensuremath{\gamma}}‐ray binaries}.
\newblock {\em Monthly Notices of the Royal Astronomical Society: Letters},
  418(1):L49--L53, 11 2011.

\bibitem{casares2005}
J.~Casares et~al.
\newblock Orbital parameters of the microquasar lsi +61 303.
\newblock {\em Monthly Notices of The Royal Astronomical Society - MON NOTIC
  ROY ASTRON SOC}, 360, 04 2005.

\bibitem{chang2016}
Z.~Chang et~al.
\newblock {Investigation of the energy dependence of the orbital light curve in
  LS 5039}.
\newblock {\em Monthly Notices of the Royal Astronomical Society},
  463(1):495--501, 08 2016.

\bibitem{distance}
D.~A. {Frail} and R.~M. {Hjellming}.
\newblock {Distance and Total Column Density to the Periodic Radio Star LSI+61
  303}.
\newblock {\em Astron. J.}, 101:2126, June 1991.

\bibitem{gregory2002}
P.~C. Gregory.
\newblock Bayesian analysis of radio observations of the be x-ray binary {LS} i
  61\textdegree 303.
\newblock {\em The Astrophysical Journal}, 575(1):427--434, aug 2002.

\bibitem{cosb}
J.~{Hermsen} et~al.
\newblock {New high energy {\ensuremath{\gamma}}-ray sources observed by COS
  B}.
\newblock {\em Nature}, 269:494–495, October 1977.

\bibitem{huber}
D.~{Huber} et~al.
\newblock {Relativistic fluid modelling of gamma-ray binaries. I. The model}.
\newblock {\em Astron. $\&$ Astrophys.}, 646:A91, February 2021.

\bibitem{bdt}
Maria {Krause}, Elisa {Pueschel}, and Gernot {Maier}.
\newblock {Improved {\ensuremath{\gamma}}/hadron separation for the detection
  of faint {\ensuremath{\gamma}}-ray sources using boosted decision trees}.
\newblock {\em Astropart. Phys.}, 89:1--9, March 2017.

\bibitem{kravtsov2020}
V.~{Kravtsov} et~al.
\newblock {Orbital variability of the optical linear polarization of the
  {\ensuremath{\gamma}}-ray binary LS I +61{\textdegree} 303 and new
  constraints on the orbital parameters}.
\newblock {\em Astron. $\&$ Astrophy.}, 643:A170, November 2020.

\bibitem{lima}
T.~P. {Li} and Y.~Q. {Ma}.
\newblock {Analysis methods for results in gamma-ray astronomy.}
\newblock {\em Astrophys. J.}, 272:317--324, September 1983.

\bibitem{eventdisplay}
G.~{Maier} and J.~{Holder}.
\newblock {Eventdisplay: An Analysis and Reconstruction Package for
  Ground-based Gamma-ray Astronomy}.
\newblock In {\em 35th International Cosmic Ray Conference (ICRC2017)}, volume
  301 of {\em International Cosmic Ray Conference}, page 747, January 2017.

\bibitem{mariaud}
C.~{Mariaud} et~al.
\newblock {VHE observations of the gamma-ray binary system LS 5039 with
  H.E.S.S}.
\newblock {\em arXiv e-prints}, page arXiv:1509.05791, September 2015.

\bibitem{massi2016}
{Massi, M.} and {Torricelli-Ciamponi, G.}
\newblock Origin of the long-term modulation of radio emission of ls i
  +61\mbox{$^\circ$}303.
\newblock {\em A$\&$A}, 585:A123, 2016.

\bibitem{atel_fast}
Shan-Shan {Weng}, ZhiChen {Pan}, Lei {Qian}, Peng {Jiang}, Ming-Yu {Ge},
  Jing-Zhi {Yan}, and Qing-Zhong {Liu}.
\newblock {FAST Detected A Transient Periodic Signal In The Direction of LS I
  +61 303}.
\newblock {\em The Astronomer's Telegram}, 14297:1, January 2021.

\bibitem{zimmermann2015}
L.~{Zimmermann}, L.~{Fuhrmann}, and M.~{Massi}.
\newblock {The broad-band radio spectrum of LS I +61{\textdegree}303 in
  outburst}.
\newblock {\em A$\&$A}, 580:L2, August 2015.

\end{thebibliography}



\clearpage \section*{Full Authors List: \Coll\ Collaboration}

\scriptsize
 \noindent
 C.~B.~Adams$^{1}$,
 A.~Archer$^{2}$,
 W.~Benbow$^{3}$,
 A.~Brill$^{1}$,
 J.~H.~Buckley$^{4}$,
 M.~Capasso$^{5}$,
 J.~L.~Christiansen$^{6}$,
 A.~J.~Chromey$^{7}$,
 M.~Errando$^{4}$,
 A.~Falcone$^{8}$,
 K.~A.~Farrell$^{9}$,
 Q.~Feng$^{5}$,
 G.~M.~Foote$^{10}$,
 L.~Fortson$^{11}$,
 A.~Furniss$^{12}$,
 A.~Gent$^{13}$,
 G.~H.~Gillanders$^{14}$,
 C.~Giuri$^{15}$,
 O.~Gueta$^{15}$,
 D.~Hanna$^{16}$,
 O.~Hervet$^{17}$,
 J.~Holder$^{10}$,
 B.~Hona$^{18}$,
 T.~B.~Humensky$^{1}$,
 W.~Jin$^{19}$,
 P.~Kaaret$^{20}$,
 M.~Kertzman$^{2}$,
 D.~Kieda$^{18}$,
 T.~K.~Kleiner$^{15}$,
 S.~Kumar$^{16}$,
 M.~J.~Lang$^{14}$,
 M.~Lundy$^{16}$,
 G.~Maier$^{15}$,
 C.~E~McGrath$^{9}$,
 P.~Moriarty$^{14}$,
 R.~Mukherjee$^{5}$,
 D.~Nieto$^{21}$,
 M.~Nievas-Rosillo$^{15}$,
 S.~O'Brien$^{16}$,
 R.~A.~Ong$^{22}$,
 A.~N.~Otte$^{13}$,
 S.~R. Patel$^{15}$,
 S.~Patel$^{20}$,
 K.~Pfrang$^{15}$,
 M.~Pohl$^{23,15}$,
 R.~R.~Prado$^{15}$,
 E.~Pueschel$^{15}$,
 J.~Quinn$^{9}$,
 K.~Ragan$^{16}$,
 P.~T.~Reynolds$^{24}$,
 D.~Ribeiro$^{1}$,
 E.~Roache$^{3}$,
 J.~L.~Ryan$^{22}$,
 I.~Sadeh$^{15}$,
 M.~Santander$^{19}$,
 G.~H.~Sembroski$^{25}$,
 R.~Shang$^{22}$,
 D.~Tak$^{15}$,
 V.~V.~Vassiliev$^{22}$,
 A.~Weinstein$^{7}$,
 D.~A.~Williams$^{17}$,
 and
 T.~J.~Williamson$^{10}$\\
 \noindent
 $^{1}${Physics Department, Columbia University, New York, NY 10027, USA}
 $^{2}${Department of Physics and Astronomy, DePauw University, Greencastle, IN 46135-0037, USA}
 $^{3}${Center for Astrophysics $|$ Harvard \& Smithsonian, Cambridge, MA 02138, USA}
 $^{4}${Department of Physics, Washington University, St. Louis, MO 63130, USA}
 $^{5}${Department of Physics and Astronomy, Barnard College, Columbia University, NY 10027, USA}
 $^{6}${Physics Department, California Polytechnic State University, San Luis Obispo, CA 94307, USA}
 $^{7}${Department of Physics and Astronomy, Iowa State University, Ames, IA 50011, USA}
 $^{8}${Department of Astronomy and Astrophysics, 525 Davey Lab, Pennsylvania State University, University Park, PA 16802, USA}
 $^{9}${School of Physics, University College Dublin, Belfield, Dublin 4, Ireland}
 $^{10}${Department of Physics and Astronomy and the Bartol Research Institute, University of Delaware, Newark, DE 19716, USA}
 $^{11}${School of Physics and Astronomy, University of Minnesota, Minneapolis, MN 55455, USA}
 $^{12}${Department of Physics, California State University - East Bay, Hayward, CA 94542, USA}
 $^{13}${School of Physics and Center for Relativistic Astrophysics, Georgia Institute of Technology, 837 State Street NW, Atlanta, GA 30332-0430}
 $^{14}${School of Physics, National University of Ireland Galway, University Road, Galway, Ireland}
 $^{15}${DESY, Platanenallee 6, 15738 Zeuthen, Germany}
 $^{16}${Physics Department, McGill University, Montreal, QC H3A 2T8, Canada}
 $^{17}${Santa Cruz Institute for Particle Physics and Department of Physics, University of California, Santa Cruz, CA 95064, USA}
 $^{18}${Department of Physics and Astronomy, University of Utah, Salt Lake City, UT 84112, USA}
 $^{19}${Department of Physics and Astronomy, University of Alabama, Tuscaloosa, AL 35487, USA}
 $^{20}${Department of Physics and Astronomy, University of Iowa, Van Allen Hall, Iowa City, IA 52242, USA}
 $^{21}${Institute of Particle and Cosmos Physics, Universidad Complutense de Madrid, 28040 Madrid, Spain}
 $^{22}${Department of Physics and Astronomy, University of California, Los Angeles, CA 90095, USA}
 $^{23}${Institute of Physics and Astronomy, University of Potsdam, 14476 Potsdam-Golm, Germany}
 $^{24}${Department of Physical Sciences, Munster Technological University, Bishopstown, Cork, T12 P928, Ireland}
 $^{25}${Department of Physics and Astronomy, Purdue University, West Lafayette, IN 47907, USA}

%
%
%

\end{document}